\documentclass[aps,prd,superscriptaddress,twocolumn,tightenlines,balancelastpage,floatfix]{revtex4}
\input epsf
\usepackage{amsmath,amssymb,subfigure}
\usepackage{graphicx}
\usepackage{epsfig}
\usepackage{color}
\usepackage{ulem}

\usepackage{epstopdf}
\usepackage{draftcopy}

\newcommand{\be}{\begin{equation}}
\newcommand{\ee}{\end{equation}}

\begin{document}

%%%%%%%%%%%%%%%%%%%%%%%%%%%%%%%%%%%%%%%%%%%%%%%%%%%%%%%%%%%%%%%%%%%%%%%%%%%%%%%%%%%%%%%%%%%%  

\title{A Multi-Parameter Investigation of Gravitational Slip}

\author{Scott F. Daniel\footnote{scott.f.daniel@dartmouth.edu}}
\affiliation{Department of Physics and Astronomy, Dartmouth College, 
Hanover, NH 03755 USA}
\author{Robert R. Caldwell}
\affiliation{Department of Physics and Astronomy, Dartmouth College, 
Hanover, NH 03755 USA}
\author{Asantha Cooray}
\affiliation{Department of Physics and Astronomy, University of California, 
Irvine, CA 92697 USA}
\author{Paolo Serra}
\affiliation{Department of Physics and Astronomy, University of California, 
Irvine, CA 92697 USA}
\author{Alessandro Melchiorri}
\affiliation{Physics Department and Sezione INFN, University of Rome,
``La Sapienza,'' P.le Aldo Moro 2, 00185 Rome, Italy}
 
\date{\today}

\begin{abstract}

A detailed analysis of gravitational slip, a new post-general relativity cosmological
parameter characterizing the degree of departure of the laws of gravitation from general
relativity on cosmological scales, is presented. This phenomenological approach assumes that
cosmic acceleration is due to new gravitational effects; the amount of spacetime curvature
produced per unit mass is changed in such a way that a Universe containing only matter and
radiation begins to accelerate as if under the influence of a cosmological constant. Changes
in the law of gravitation are further manifest in the behavior of the inhomogeneous
gravitational field, as reflected in the cosmic microwave background, weak lensing, and
evolution of large-scale structure. The new parameter, $\varpi_0$, is naively expected to be
of order unity. However, a multiparameter analysis, allowing for variation of all the
standard cosmological parameters, finds that $\varpi_0 = 0.09{}^{+0.74}_{-0.59}\, (2\sigma)$
where $\varpi_0=0$ corresponds to a $\Lambda$CDM universe under general relativity. Future
probes of the cosmic microwave background (Planck) and large-scale structure (Euclid)
may improve the limits by a factor of four.

\end{abstract}

\maketitle

%%%%%%%%%%%%%%%%%%%%%%%%%%%%%%%%%%%%%%%%%%%%%%%%%%%
\section{Introduction}
 
Cosmic acceleration \cite{Riess:1998cb,Perlmutter:1998np} can be caused by
new fluids, new theories of gravity, or some admixture of both \cite{Uzan:2006mf}. This
uncertainty places a premium on descriptions of the so-called ``dark physics'' which remain
useful across different models and in spite of varying assumptions.  In the case of new
fluids (dark energy), the literature chooses to speak in terms of the equation of state $w$
and its derivative \cite{Albrecht:2006um}. In the case of new gravitational physics, the
model-independent {\it lingua franca} is the relationship between the Newtonian ($\psi$) and
longitudinal ($\phi$) gravitational potentials. The potentials, implicitly defined through
the perturbed Robertson-Walker metric 
\begin{equation}
ds^2=a^2[-(1+2\psi)d\tau^2+(1-2\phi)d\vec{x}^2],
\end{equation}
are most familiar for their roles in Newton's equation, $\ddot{\vec x}=-\vec{\nabla}\psi$,
and the Poisson equation, $\nabla^2\phi=4\pi G a^2\delta\rho$ under general relativity (GR).

The gravitational potentials are equal in the presence of non-relativistic stress-energy
under GR. Alternate theories of gravity
make no such guarantee. Scalar-tensor
\cite{Carroll:2004st,Schimd:2004nq} and $f(R)$ theories
\cite{Capozziello:2003tk,Acquaviva:2004fv,Zhang:2005vt}, braneworld scenarios
such as Dvali-Gabadadze-Porrati gravity \cite{Dvali:2000hr,Lue:2005ya,Song:2006jk}, and massive gravity
\cite{Dubovsky:2004sg,Bebronne:2007qh}  all predict a systematic difference or ``slip", so
that $\phi\ne\psi$ in the presence of non-relativistic stress-energy. 
Efforts to develop a
parametrized-post-Friedmannian (PPF) framework to phenomenologically describe this behavior
are just as prolific: Refs.~
\cite{Bertschinger:2006aw,Caldwell:2007cw,Zhang:2007nk,Hu:2007pj,Amendola:2007rr,Jain:2007yk,Zhang:2008ba,Daniel:2008et,Bertschinger:2008zb,Hu:2008zd}
all offer parametrizations quantifying the departure from $\phi=\psi$ due to new
gravitational effects. We choose to work with the parametrization proposed in
Ref.~\cite{Caldwell:2007cw}: 
\begin{eqnarray}
\psi&=&[1+\varpi(z)]\phi\label{parametrization}\\
\varpi(z) &=& \varpi_0 (1+z)^{-3}.\label{varpieqn}
\end{eqnarray}
We assume the existence of a theory of gravitation that leads to an expansion history that
is indistinguishable from that produced by a spatially-flat, $\Lambda$CDM scenario with
density parameters $\Omega_m$ and $\Omega_\Lambda = 1-\Omega_m$.  This assumption is not
essential, but it allows our analysis to focus solely on PPF effects. Our naive expectation
is that $\varpi \simeq \Omega_\Lambda/\Omega_m$ by today. [Note that we have changed our
notation, having previously defined $\varpi(z) = \varpi_0 (\Omega_\Lambda/\Omega_m)
(1+z)^{-3}$.]

The departure from GR kicks in only when the cosmic expansion begins to accelerate. Daniel
{\it et al.} \cite{Daniel:2008et} (hereafter DCCM) discuss the compatibility with other
parametrizations (especially that of Ref.~\cite{Bertschinger:2008zb}) and compare the
implications of $\varpi_0 \neq 0$ to data from the Wilkinson Microwave Anisotropy Probe
(WMAP) \cite{wmap3data}, the Canada-France-Hawaii Telescope Legacy Survey (CFHTLS)
\cite{Fu:2007qq}, and various galaxy surveys
\cite{Gaztanaga:2004sk,Giannantonio:2006du,Cabre:2006qm}. We expand upon their analysis in
this work by performing a full likelihood analysis of the cosmological parameter space.

The previous work by DCCM considered the effects of modified gravity on cosmological
perturbations in a one-parameter context: i.e., ``how does the the new (modified gravity)
parameter affect cosmological data when all  other parameters are held fixed (at the WMAP 3
year maximum likelihood values)?'' They used a modified version of the Boltzmann code
CMBfast \cite{Seljak:1996is} to evaluate the effect of $\varpi_0$  on the cosmic microwave
background (CMB) anisotropy,  matter power spectrum, weak lensing convergence
correlation function, and galaxy-CMB cross-correlation power spectrum. While this analysis
was useful for testing for the existence of PPF effects, the results glossed
over degeneracies that exist between $\varpi_0$ and traditional cosmological parameters.
DCCM's Figure 9 already demonstrates a potential degeneracy between $\varpi_0$ and
$\sigma_8$. Identifying further degeneracies and more rigorously motivating the possiblity
of non-zero $\varpi_0$ requires analysis across the full cosmological parameter space. 

In the following, we present the results of a likelihood analysis based on a Monte Carlo
Markov chain sampling of the space of cosmological parameters. The parameters, $\{\Omega_b
h^2, \Omega_c h^2, \theta, \tau_\text{ri}, n_s, A_s, A_{\text{SZ}}, \varpi_0\}$,   are respectively  the baryon
density, cold dark matter density, the ratio of the sound horizon to the angular diameter
distance, the optical depth to last scattering, the scalar spectral index, the amplitude
of the primordial curvature perturbations, and a normalization parameter for the SZ effect.
These are the standard parameters in the
convention used by the publicly-available code CosmoMC \cite{CosmoMCreadme}.

We generate our Markov Chains using CosmoMC \cite{Lewis:1999bs,Lewis:2002ah,CosmoMC_notes} 
with modules added to calculate likelihoods based on the weak lensing
\cite{Massey:2007gh,Lesgourgues:2007te} and galaxy-CMB cross-correlation spectra
\cite{Ho:2008bz}. The CMB data and likelihood code comes from the WMAP team's 5-year release
\cite{Dunkley:2008ie}. Supernova data comes from the Union data set produced by the
Supernova Cosmology Project \cite{Kowalski:2008ez}. The weak lensing data comes from the
CFHTLS weak lensing survey \cite{Fu:2007qq,Kilbinger:2008gk}.

To help understand our results, we present a closed system of ordinary differential 
equations describing the evolution of $\phi$ and the matter overdensity $\delta$ under
$\varpi_0\ne0$. These results imply a correction to the Poisson equation that was neglected
in DCCM. Section \ref{quadrupole} presents these equations and uses them to describe the
dependence of the large-angle CMB anisotropy on $\varpi_0$. Section \ref{code} discusses the
modifications made to the public CosmoMC codes to implement Eq.~(\ref{parametrization}).
Section \ref{results} presents the likelihood contours found from our Markov chains. Section
\ref{forecasts} makes an attempt at forecasting the results of
future experiments.  We conclude in Section \ref{conclusions}.

%%%%%%%%%%
\section{Evolution of Perturbations}
\label{quadrupole}

The procedure for evolving the matter and metric perturbations is as follows.  We assume
that the perturbed stress-energy tensors for all matter and radiation are conserved
independently of the theory of gravitation:
\begin{equation}
\nabla_\mu T^{\mu\nu}=0.
\label{tmunueqn}
\end{equation} 
We next impose the relationship given by Eq.~(\ref{parametrization}) between potentials
$\phi$ and $\psi$, which upon translation into synchronous gauge implies an evolution
equation for the metric variable $\alpha \equiv (\dot h + 6 \dot\eta)/2 k^2$:
\begin{equation}
\dot{\alpha}=-(2+\varpi)\mathcal{H}\alpha+(1+\varpi)\eta  
 -12\pi G a^2(\bar{\rho}+\bar{p})\sigma/k^2. \label{alphadot}
\end{equation}
Here, a dot indicates the derivative with respect to conformal time, $h$ and $\eta$ are the
synchronous-gauge metric perturbations, $\mathcal{H} = \dot a/a$ is the conformal-time
Hubble parameter, and $\sigma$ is the shear in a fluid with mean density $\bar\rho$ and
pressure $\bar p$. (We use the same notation as Ref.~\cite{Ma:1995ey}.)  We further assume
that there is no preferred reference frame introduced by the new gravitational effects;
there is no  ``dark fluid" momentum flux or velocity relative to the dark matter and baryon
cosmic rest frame. This condition is imposed by enforcing the same perturbed time-space
equation as in GR, 
\begin{equation}
k^2\dot\eta = 4\pi G a^2(\bar\rho+\bar p)\theta ,
\label{etadot}
\end{equation}
where $\theta$ is the divergence of the velocity field in a fluid with mean density
$\bar\rho$ and pressure $\bar p$. Satisfying this equation automatically means that
Bertschinger's consistency condition, that long-wavelength curvature perturbations should
evolve like separate Robertston-Walker spacetimes, is satisfied \cite{Bertschinger:2006aw}. The model of $\varpi(z)$
plus the three Eqs.~(\ref{tmunueqn}-\ref{etadot}) close the system of equations. (See
Refs.~\cite{Caldwell:2007cw,Daniel:2008et} for further details.) 
In order to study the late-time behavior of the system of equations, we
may neglect the shear and velocity perturbations and express the evolution equations 
in conformal-Newtonian/longitudinal gauge as
\begin{eqnarray}
\ddot{\phi}&=&-(3+\varpi)\mathcal{H}\dot{\phi}-\dot{\varpi}\mathcal{H}\phi 
-(1+\varpi)(\mathcal{H}^2+2\dot{\mathcal{H}})\phi,\quad\label{phidoteqn}\\
\dot{\delta}&=&3\dot{\phi}-
\left(\frac{k}{\mathcal{H}}\right)^2
\frac{\dot{\phi}+(1+\varpi)\mathcal{H}\phi}{1-\dot{\mathcal{H}}/\mathcal{H}^2 }\label{deltadoteqn}
\end{eqnarray}
where $\delta$ is the matter density contrast.

\begin{figure}[!t]
\includegraphics[scale=0.35]{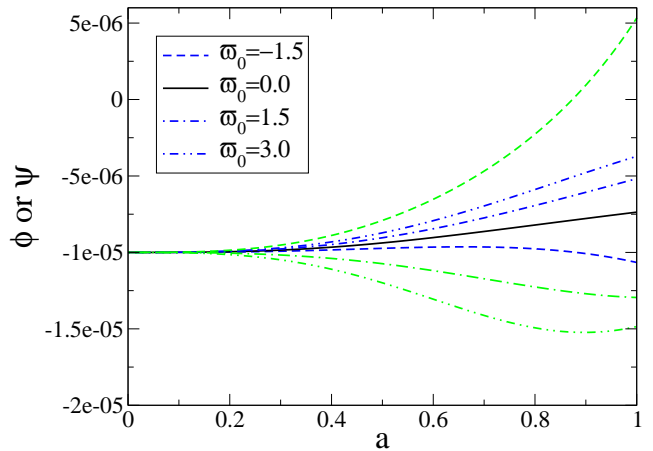}
\caption{The potentials $\phi$ and $\psi$ are shown versus the scale factor, for different
values of $\varpi_0$. The blue, dark curves are $\phi$, whereas the green, light curves are
$\psi$. Note that they behave oppositely; when $\phi$ becomes shallower, $\psi$ becomes
deeper, and {\it vice versa}.}
\label{phifig}%
\end{figure}

\begin{figure}[!t]
\includegraphics[scale=0.35]{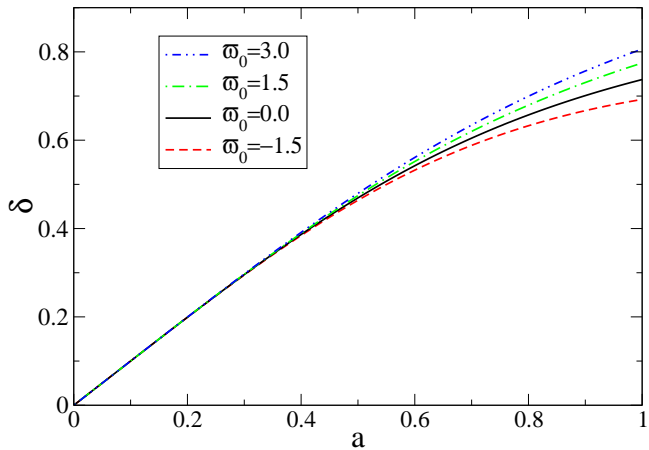}
\caption{The matter density contrast is shown versus the scale factor, for different values
of $\varpi_0$. The time evolution is obtained by 
integrating Eqs.~(\ref{phidoteqn}) and (\ref{deltadoteqn}), with 
initial conditions $\phi=-10^{-5}$, $\dot{\phi}=0.0$ for $k=0.01 \text{Mpc}^{-1}$.
Positive (negative) values of $\varpi_0$ enhance (slow) the growth of density perturbations.}
\label{deltafig}%
\end{figure}

\begin{figure}[!t]
\includegraphics[scale=0.35]{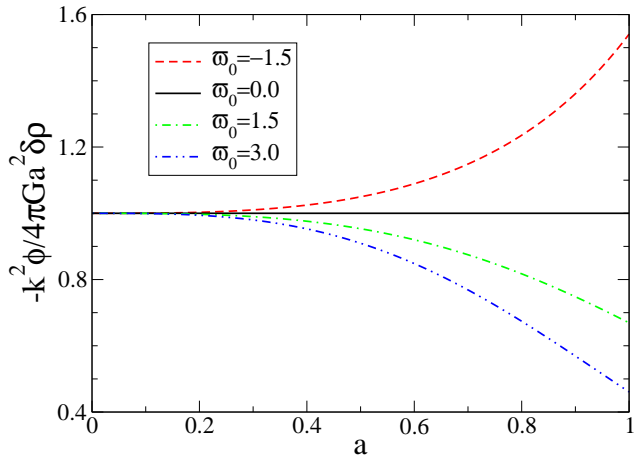}
\caption{The degree of deviation from the Poisson equation versus scale factor is shown for
different values of $\varpi_0$. Because $\varpi$ is scale-independent, so too is  the ratio
$-k^2 \phi/(4 \pi G a^2 \delta\rho)$. For positive (negative) $\varpi_0$, a given $\phi$
corresponds to a larger (smaller) density contrast than in GR.}
\label{poissonfig}%
\end{figure}

Consider the behavior of an overdense region, $\delta >0$, as it evolves from early times
when GR is valid to late times when new gravitational effects characterized by $\varpi$ 
become important. At early times, when the Poisson equation is valid, $\phi<0$ for the
overdensity. While the expansion is matter-dominated, the potential remains static. However,
at late times, with the onset of cosmic acceleration, the potential begins to evolve. In the
case of GR, $\dot{\phi}>0$ so the potential is stretched shallower. The density contrast
$\delta$ continues to grow via gravitational instability, although the rate of growth is
slowed.
The evolution of $\phi$ can be understood in terms of a competition between
the expansion diluting the matter density and stretching $\phi$ shallower, 
and the accretion of matter sourcing and deepening $\phi$.
In GR, the accelerated
expansion upsets the balance in favor of dilution, so that $\phi$ becomes shallower and
$\delta$ grows more slowly. When $\varpi_0 \neq 0$, the competition between effects changes.
Numerically integrating Eqs.~(\ref{phidoteqn}) and (\ref{deltadoteqn}), we find that
$\varpi_0>0$ causes $\phi$ to become even shallower, yet the density contrast grows faster,
as illustrated in Figs.~\ref{phifig} and \ref{deltafig}. This seems counter-intuitive, since
the shallower potential should provide weaker attraction for the accretion of surrounding
matter. In the case $\varpi_0<0$, the potential $\phi$ becomes more negative or deeper, and
the density contrast grows more slowly. Likewise, the deeper potential should provide
greater attraction. But here the difference between $\phi$ and $\psi$ is important. As seen
in Fig.~\ref{phifig}, the potential $\phi$ grows shallower (deeper) for $\varpi_0 >0$
($<0$). However, the potential responsible for acceleration $\psi = (1+\varpi)\phi$ 
behaves oppositely, becoming
deeper (shallower). Hence, the competition swings in favor of increased clustering over
dilution by the expansion.

The new behavior of $\phi$ and $\delta$ implies a correction to
the Poisson equation. As seen in Fig.~\ref{poissonfig}, for $\varpi_0>0$ ($<0$), the density
contrast grows more (less) rapidly and the potential $\phi$ becomes shallower (deeper), so
that the ratio 
\begin{equation}
\Gamma \equiv -k^2 \phi / (4 \pi G a^2 \delta\rho) 
\label{curlygeqn}
\end{equation} 
grows smaller (larger). This suggests that we can restore the Poisson equation by
introducing a time-dependent gravitational constant $G_{eff} = G\, \Gamma$, whence $-k^2
\phi = 4 \pi G_{eff} a^2 \delta\rho$. Note that $G_{eff}$ is not a free function, but is
determined by Eqs.~(\ref{tmunueqn}-\ref{etadot}). Because we have chosen $\varpi$ to be
scale-independent, $G_{eff}$ is too. A different strategy,  whereby the time- and
space-dependence of $G_{eff}$ is imposed separately \cite{Jain:2007yk}, 
will not necessarily satisfy Eqs.~(\ref{tmunueqn}-\ref{etadot}).

\begin{figure}[!t]
\includegraphics[scale=0.35]{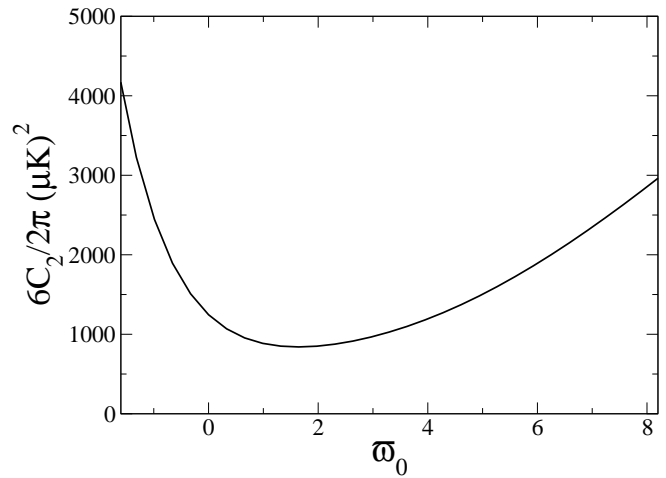}
\caption{The CMB quadrupole moment is shown versus $\varpi_0$. As explained in the text, the
quadratic dependence can be understood in terms of the influence of $\varpi_0$ on the ISW
effect.(Reproduced from DCCM with our new normalization Eq.~(\ref{varpieqn}).)}
\label{quadrupolefig}%
\end{figure}

\begin{figure}[!t]
\includegraphics[scale=0.35]{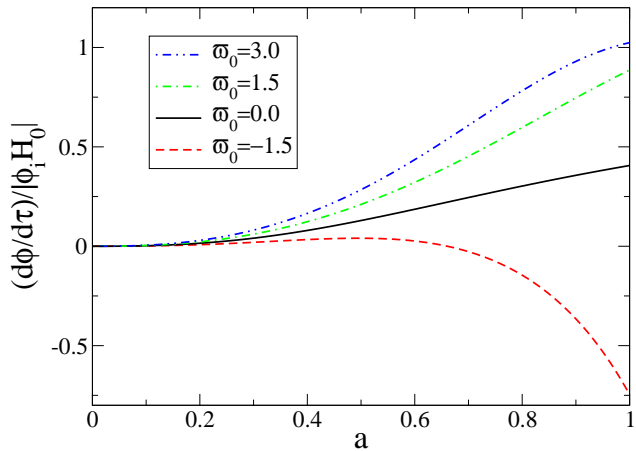}
\caption{The conformal time derivative of the gravitational potential $\phi$ is shown versus
scale factor, for different values of $\varpi_0$. Initial conditions are the same as 
in Fig. ~\protect{\ref{deltafig}}. The
potential well is decaying when $\frac{d\phi}{d\tau}/|\phi_i H_0|>0$, and is deepening
when negative.}
\label{phidotfig}%
\end{figure}

We can use this new understanding to explain the curious behavior of the large-angular scale
CMB anisotropy spectrum. The effect of $\varpi_0\ne0$ on the low $l$ moments of the CMB
anisotropy is not monotonic, as seen in Fig.~\ref{quadrupolefig}. The cause is the
suppression of the integrated Sachs-Wolfe effect (ISW) at $\varpi_0\simeq 1$.  If the
gravitational potentials in the Universe are evolving with time, CMB photons will lose less
(more) energy climbing out of potential wells than they gained falling in, resulting in a
net blue (red) shift as the potentials shrink (grow). This is the ISW effect whereby
time-evolving gravitational potentials contribute to the moments of the photon distribution
function, $\Theta_l(k,\,\eta)$, via
\begin{equation}
\label{phidotpsidotbessel}
\int_0^{\tau_0} d\tau(\dot{\phi}(k,\tau)+\dot{\psi}(k,\tau))
j_l(k(\tau_0-\tau))\exp[-\tau_\text{ri}(z)].
\end{equation}
(See equation 8.55 of Ref.~\cite{Dodelson:2003ft}.) Here $j_l$ is a spherical Bessel
function of the first kind, $\tau$ is the conformal time, $\tau_0$ is the conformal time at
$z=0$, and $\tau_\text{ri}(z)$ is the optical depth to redshift $z$. The strength of the ISW effect
is determined by the sum $\dot\phi + \dot\psi$, which, using
Eqs.~(\ref{parametrization}-\ref{varpieqn}), is given by
\begin{eqnarray}
\dot{\phi}+\dot{\psi}&=&\dot{\phi}(2+\varpi)+\phi\dot{\varpi}\nonumber\\
&=&\dot{\phi}(2+\varpi)+3\phi\mathcal{H}\varpi. \label{quadisw}
\end{eqnarray}
Again consider the evolution of an overdensity $\delta>0$ with $\phi <0$. In GR, the sum is
positive, $\dot\phi + \dot\psi>0$. When $\varpi_0<0$, the second term in 
Eq.~(\ref{quadisw}) is always positive. The first term is generally subdominant, since
$|\dot\phi| < \mathcal{H} \phi$, as can be inferred from Fig.~\ref{phidotfig}. Therefore
$\varpi_0<0$ enhances the ISW effect.  When $\varpi_0 >0$, there is a competition between
the first and second terms; the first term is positive, whereas the second term is negative,
The first term always wins, but at some intermediate value of $\varpi_0$ the two terms
nearly cancel, thereby suppressing the ISW effect relative to the case with $\varpi_0=0$.
This explains the dip in the quadrupole moment versus $\varpi_0$, seen in
Fig.~\ref{quadrupolefig}.

%%%%%%%%%%%%%%%%%%%%%%%%%%%
\section{Implementation}
\label{code}

The modifications of the Monte Carlo Markov chain software CosmoMC to allow for $\varpi_0\ne
0$ proceed almost identically to the modifications made to CMBfast for DCCM, with a few
differences. 
To compare the predictions of our model with weak lensing data we adapt the weak lensing 
module provided by Refs.~\cite{Massey:2007gh,Lesgourgues:2007te}. We modify it
to assess the likelihood  in terms of the variance of the aperture mass (Eq.~5 of
\cite{Fu:2007qq}) with a full covariance matrix \cite{CFHTLSdata}. 
Because we probe weak-lensing at non-linear scales, we calculate
the power spectrum of the lensing potential by extrapolating the linear matter power spectrum
$P_\delta$ to non-linear scales and using the relationship (\ref{curlygeqn})
between the matter 
overdensity $\delta$ and the gravitational potential $\phi$ to find the non-linear $P_\phi$.
Whereas CMBfast calculates the non-linear matter power spectrum from the
phenomenological fit of Peacock and Dodds \cite{Peacock:1996ci}, CosmoMC (having been built
around the  code CAMB \cite{Lewis:1999bs}) uses Smith {\it et al.}'s fit \cite{Smith:2002dz}
(see their Appendix C). Smith {\it et al.} express their fit as a non-trivial function of
the linear power spectrum and $\Omega_m$. This function assumes the $\Lambda$CDM
relationship between $\Omega_m$ and perturbation growth. Gravitational slip alters this
relationship, as discussed above in section \ref{quadrupole}. Therefore, to adapt Smith {\it
et al.}'s fit to the case $\varpi_0\neq 0$, we use the 
phenomenological relationship (DCCM equation 24)
\begin{equation}
\label{omegamfit}
\Omega_{m}|_{\varpi_0=0}=\Omega_{m}|_{\varpi_0\ne0}+0.13\varpi_0\frac{\Omega_m}{\Omega_\Lambda}
\end{equation}
to find a $\varpi_0=0$, $\Lambda$CDM model with a similar growth history to our 
$\varpi_0\ne 0$ model and use that value of $\Omega_{m}|_{\varpi_0=0}$ in Smith {\it et
al.}'s equations (C18). Eq.~(\ref{omegamfit}) breaks down for $\Omega_{m}|_{\varpi_0\ne0}
\le 0.15$, but this region of parameter space is excluded to at least $2\sigma$ (see Figure
\ref{omgfig}). A second CosmoMC run with a more accurate fitting function yielded identical
results to those obtained using Eq.~(\ref{omegamfit}).

%%%%%%%%%%
This is not a precise method for determining the non-linear power spectrum
in the presence of gravitational slip.  Precision would require
examination of N-body simulations which, unfortunately, implies
assumptions about what alternative theory of gravity we are constraining.
Recently, much work has been done attempting to calculate the non-linear
power spectrum directly, without the aid of an N-body simulation.  Crocce and
Scoccimarro propose to expand the non-linear power spectrum
as a Taylor-like sum
\begin{equation}
\label{psum}
P_\delta=\sum_iP_\delta^{(i)}
\end{equation}
where the different orders of $P_\delta$ are derived from a diagrammatic scheme
similar to Feynman diagrams \cite{Crocce:2005xy}.  
They find that the resulting sum (\ref{psum}) is much better behaved than
results derived from perturbation theory
(see their Figure 1).
Matarrese and Pietroni \cite{Matarrese:2007wc}
use the formalism of renormalization group theory to
derive a generating functional for the different orders of $P_\delta$.  Taruya
and Hiramatsu adapt methods from the statistical studies of fluid instabilities
to separate out and solve for the cross-mode interactions in $\tilde\delta$
\cite{Taruya:2007xy}.  
All of these methods yield better agreement with the results of N-body
simulations than standard perturbation theory in the case of $\varpi=0$ 
(see Figure 2 of 
Ref.~\cite{Crocce:2007dt}, Figure 8 of Ref.~\cite{Matarrese:2007wc}, and Figure
3 of Ref.~\cite{Hiramatsu:2009ki}).
Work has already begun adapting them to alternative gravity theories.
In Ref.~\cite{Koyama:2009me}, Koyama, Taruya, and
Hiramatsu extend the method of Ref.~\cite{Taruya:2007xy} to include $f(R)$
and DGP gravity theories by assuming that they can be approximated 
with a Brans-Dicke scalar-tensor theory on sub-horizon scales.  Hiramatsu and
Taruya \cite{Hiramatsu:2009ki} 
also try to encompass modified gravity theories by parametrizing them in
terms of their implied effective Newton's constant $G_\text{eff}=\Gamma G$ (see
equation \ref{poissoncorrection} of the present work).  
Following their lead, it
should be possible to adapt the non-linear power spectrum calculations of
Ref.~\cite{Taruya:2007xy} -- or even \cite{Crocce:2005xy} and
\cite{Matarrese:2007wc} -- to account for model-independent gravitational slip.
Such a calculation is beyond the scope of this work.  Given the relatively
well-behaved regions of parameter space allowed by 
experiments (see Section \ref{results} below),
we do not expect this limitation to significantly influence our findings.

%%%%%%%%
To incorporate the
galaxy-CMB cross-correlation, we use the module written by  Ho {\it et al.}
\cite{Ho:2008bz}. Modifications for $\varpi_0\ne 0$ enter as modifications to the
$\phi+\psi$ power spectrum (see section II of \cite{Ho:2008bz}), 
\begin{equation}
\label{poissoncorrection}
P_{\phi+\psi} =
\frac{9}{4}\Omega_{m,0}^2\left(\frac{H_0}{ck}\right)^4
\left(\frac{D_{\varpi}}{a}\right)^2
\left[(1 + \frac{1}{2}\varpi)\Gamma\right]^2 \times P_{\delta}.
\end{equation}
Note that equation (27) of DCCM neglected the factor $\Gamma$, defined in
Eq.~(\ref{curlygeqn}), to correct the Poisson equation. The corrected weak lensing statistics
show the same qualitative behavior as in Figure 10 of DCCM. However, large values
$|\varpi_0| \gg 1$ have a weaker effect on the amplitude of the convergence spectrum.

%%%%%%%%%%%%%%%%%%%%%%%
\section{Results}
\label{results}

\begin{figure}[!t]
\includegraphics[scale=0.42]{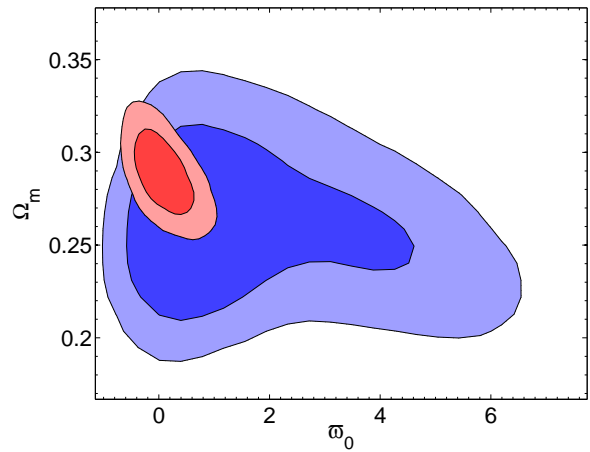}
\caption{The 68\% and 95\% likelihood contours in the $\varpi_0-\Omega_m$ parameter
space are shown. The blue contours are based on CMB data alone.
The red contours add  weak lensing, type 1a supernovae, and galaxy-CMB cross-correlation
data.}
\label{omgfig}%
\end{figure}

\begin{figure}[!t]
\includegraphics[scale=0.42]{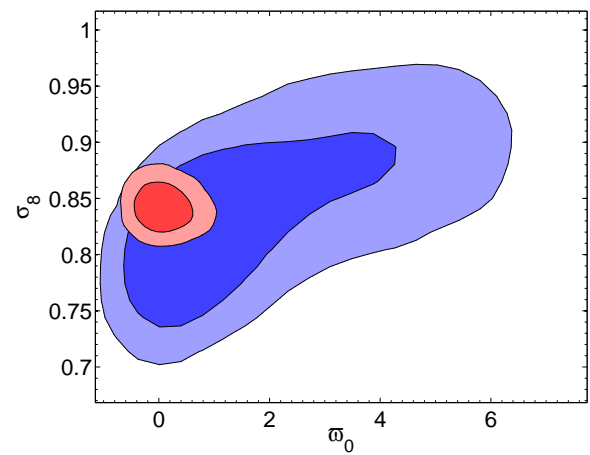}
\caption{The 68\% and 95\% likelihood contours in the $\varpi_0-\sigma_8$ parameter
space are shown. Shading is the same as in Fig.~\protect{\ref{omgfig}}. Note that the
addition of large-scale structure data breaks the degeneracy in $\varpi_0-\sigma_8$.}
\label{sigfig}%
\end{figure}

The results of our multiparameter investigation are shown in Figs.~\ref{omgfig},
\ref{sigfig}. Fig.~\ref{omgfig} shows the $68\%$ and $95\%$ contours in 
$(\Omega_m,\varpi_0)$ space marginalized over all other parameters. Fig.~\ref{sigfig}
shows the same contours in $(\sigma_8,\varpi_0)$ space. Red (smaller) likelihood contours were
generated using all available data sets (WMAP 5 year \cite{Dunkley:2008ie}, Supernova Union
\cite{Kowalski:2008ez}, CFHTLS \cite{Fu:2007qq}, and the galaxy surveys selected by
\cite{Ho:2008bz}).  Blue (larger) contours were generated using only the WMAP 5 year data.
For each set of constraints, we generated four independent Markov Chains.  We achieved
convergence by running the calculations until the statistic $|1-R|$ was much less than
unity, 
where $R$ is Gelman and Rubin's potential scale reduction factor,
defined as the ratio of the variance across all of the chains to the mean of the variance
of each individual chain evaluated for the least converged parameter.
 \cite{GelmanRubin,BrooksGelman,CosmoMCreadme}.  
 Our conclusions are three-fold:
\begin{itemize} 
\item Present cosmological data constrains gravity to agree with GR, assuming a background
evolution consistent with $\Lambda$CDM.
\item Very negative values of $\varpi_0$ are ruled out. This should not be surprising, since
a sign difference between the longitudinal and Newtonian gravitational potentials would mean
that test particles are repelled  by overdense regions. 
\item Large-scale structure data (in our case, weak lensing  and the galaxy-CMB correlation)
are critical to constraining $\varpi_0$.
\end{itemize}
The effects described in section \ref{quadrupole} mean that any CMB anisotropy spectrum can
be reasonably well approximated (modulo a normalization) by two possible values of
$\varpi_0$.  DCCM Figure 1 showed that $\varpi_0\ne 0$ has no effect on the shape of higher
$l$ multipoles within linear theory. This explains the double-peaked likelihood curve in DCCM Fig.~3 and the
broad blue contours in  Figs.~\ref{omgfig} and \ref{sigfig} in this work. Fortunately,
the effect of $\varpi_0\ne 0$ on cosmic structure is monotonic in the range of interest (as
discussed in DCCM), so that only one value of $\varpi_0$ is maximally likely for any given
realization of weak lensing and galaxy-CMB cross-correlation data, hence the smaller red
contours in Fig.~\ref{omgfig} and \ref{sigfig}. Marginalizing over all other parameters, the
WMAP 5 year data alone gives $\varpi_0=1.7^{+4.0}_{-2.0} \, (2\sigma)$. Including supernovae, weak
lensing, and the galaxy-CMB cross-correlation data improves the constraint to $\varpi_0 =
0.09{}^{+0.74}_{-0.59}\, (2\sigma)$.  Table \ref{paramtable} presents the 
marginalized $1\sigma$ limits on the other cosmological parameters of note.

\begin{table*}[!t]
\begin{tabular}{l l l l}
%add WMAP5 parameters in fourth column
parameter&\qquad$\varpi_0\ne0$&\qquad$\varpi_0=0$&\qquad WMAP 5-year\\
\hline
$\Omega_b h^2$&\qquad$0.02262^{+0.00059}_{-0.00058}$&\qquad$0.02264_{-0.00057}^{+0.00058}$&\qquad$0.02273\pm0.00062$\\
$\Omega_\text{cdm}h^2$&\qquad$0.1167\pm0.0026$&\qquad$0.1170\pm0.0016$&\qquad$0.1109\pm0.0062$\\
$\theta_s$&\qquad$1.0417_{-0.0028}^{+0.0029}$&\qquad$1.0419_{-0.0029}^{+0.0028}$&\qquad$1.0400\pm0.0029$\\
$\tau_\text{ri}$&\qquad$0.085\pm0.016$&\qquad$0.087_{-0.016}^{+0.017}$&\qquad$0.087\pm0.017$\\
$n_s$&\qquad$0.964\pm0.014$&\qquad$0.965\pm0.014$&\qquad$0.963_{-0.015}^{+0.014}$\\
%$\ln[10^{10}A_s]$&\qquad$3.206^{+0.037}_{-0.038}$&\qquad$3.207\pm0.039$\\
$\Omega_\Lambda$&\qquad$0.712\pm0.014$&\qquad$0.710_{-0.011}^{+0.012}$&\qquad$0.742\pm0.030$\\
$\sigma_8$&\qquad$0.842\pm0.014$&\qquad$0.844\pm0.015$&\qquad$0.796\pm0.036$\\
$h$&\qquad$0.696\pm0.014$&\qquad$0.695\pm0.013$&\qquad$0.719^{+0.026}_{-0.027}$
\end{tabular}
\caption{Marginalized ($1\sigma$) constraints for cosmological
parameters resulting from Monte Carlo Markov chain analysis.
The left and center columns are generated using all available data sets
(CMB, weak lensing, supernovae, and galaxy-CMB cross-correlation).
The left column is generated allows $\varpi_0$ to vary.  The center column
fixes $\varpi_0=0$.  Because our constraint on $\varpi_0$ is
consistent with $\varpi_0=0$, we find little difference 
between the two columns.  The right column shows the constraints 
reported by the WMAP team in Ref.~\cite{Dunkley:2008ie}
based on just the WMAP 5-year data.  
The principal improvements from adding supernova, weak lensing, and galaxy-CMB
cross-correlation data lie in constraining $\Omega_\Lambda$ 
(a result of adding the supernovae) and $\sigma_8$
(a result of adding weak lensing).}
\label{paramtable}
\end{table*}

%%%%%%%%%%%%%%%%%%%%
\section{Forecasts}
\label{forecasts}

\begin{figure}[t]
\includegraphics[scale=0.42]{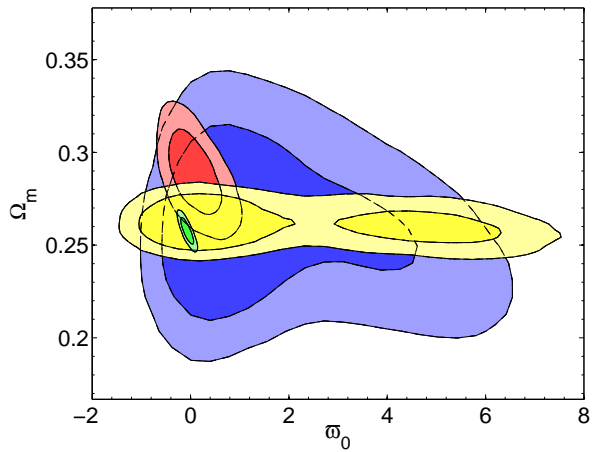}
\caption{The projected 
68\% and 95\% likelihood contours in the $\varpi_0-\Omega_m$ parameter
space are shown. The yellow contours are based on mock Planck data.
The green contours add mock weak lensing data. The underlying model is assumed to
be $\varpi_0=0$ with $\Omega_m=0.26$. The current constraints are shown for
reference.}
\label{omgfigproj}%
\end{figure}

\begin{figure}[t]
\includegraphics[scale=0.42]{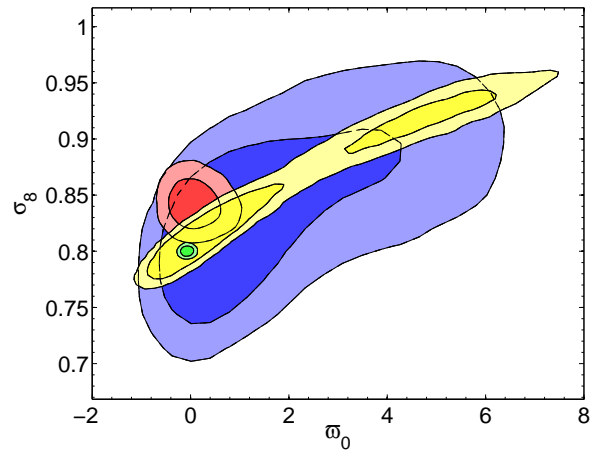}
\caption{The projected 
68\% and 95\% likelihood contours in the $\varpi_0-\sigma_8$ parameter
space are shown. Shading is the same as in Fig.~\protect{\ref{omgfigproj}}.}
\label{sigfigproj}%
\end{figure}

It is useful to ask how much better our constraints
will be under future experiments. We generate two mock data sets -- one simulating the
results  of the upcoming Planck CMB experiment, the other simulating the results of a future
weak lensing survey, modeled after the proposed ESA experiment Euclid -- and feed them into
our modified CosmoMC.  

To simulate Planck data we use a fiducial model given by the best fit parameters of WMAP
\cite{Dunkley:2008ie} with noise properties consistent with a combination of Planck
$100$-$143$-$217$ GHz channels of HFI \cite{bluebook}; in this case we fit also for
B-modes produced by
lensing of the CMB (see Ref.~\cite{Paolo})
and we use the full-sky likelihood function given in \cite{Lewis:2005tp}.

To simulate weak lensing data, we generate a mock convergence power spectrum $P_\kappa (l)$
(equation (2) of Ref.~\cite{Fu:2007qq}) corrected for alternative gravity as in
Eq.(\ref{poissoncorrection}). We generate data in bins of size 
$\Delta_l=1$ for $2\le l<100$ and $\Delta_l=40$ for $100<l<2980$. We simulate the
(1$\sigma$) errors as (Eq.~11 of Ref.~\cite{Cooray:1999rv})
\begin{equation}
\nonumber
\sigma_l=\sqrt{(2/(2l+1))/(\Delta_l
f_{\text{sky}})}(P_{\kappa}(l)+\sigma_\epsilon^2/n_\text{gal}),
\end{equation}
taking $\sigma_\epsilon=0.25$, $n_{\text{gal}}=35 \text{(arc minute)}^{-2}$ and $f_\text{sky}=0.48$,
consistent with values projected for ESA's 
Euclid experiment (Table 1 of Ref.~\cite{Kitching:2008dp}). 
These assumptions will give us a tighter constraint than if we had used SNAP/JDEM parameters,
since SNAP/JDEM has a smaller $f_\text{sky}$ by a factor of $10$ \cite{SNAP}.
We fit the redshift distribution of sources $n(z)$ from a mock data
set based on Eq.~14 of Ref.~\cite{Fu:2007qq} with parameter values taken from their
Table 1.  The 1$\sigma$ errors in our mock $n(z)$ are reduced from actual values
\cite{CFHTLSdata} by a factor of $1/\sqrt{2}$.  The likelihood relative
to the mock weak lensing data is calculated as a simple
$\chi^2$ (i.e., we assume that the covariance matrix is diagonal).  This is a
safe assumption according to \cite{Cooray:2000ry}. Figs.~\ref{omgfigproj}
and \ref{sigfigproj} show the resulting likelihood contours.

Looking at the Planck-only (yellow) contours, we see the weakness of using CMB measurements
alone to constrain $\varpi_0$, as a bimodal distribution is obtained once again. We also see
more clearly in Fig.~\ref{sigfigproj} the degeneracy between $\varpi_0$ and $\sigma_8$ as
normalization parameters (one can interpret the effect of $\varpi_0$ on $\delta$ in
Fig.~\ref{deltafig} as a renormalization of the matter power spectrum). Since weak lensing
statistics depend sensitively on the power spectrum normalization, they once again break the
degeneracy.  Marginalizing over all other parameters, the mock datasets give the constraint
$\varpi_0=-0.07{}^{+0.13}_{-0.16}\,(2\sigma)$, a factor of $\sim 4$ improvement over the
current constraint.

%%%%%%%%%%%%%%
\section{Conclusions}
\label{conclusions}

If we are justified in describing the background evolution by a $\Lambda$CDM universe, then
the results illustrated in Figs.~\ref{omgfig} and \ref{sigfig} do not appear to indicate a
significant departure from GR. 
In fact,these results conflict with our naive expectation 
that $\varpi_0 \simeq\Omega_\Lambda / \Omega_m$. However, these results allow the ratio 
of $\phi$ to $\psi$ to vary
by order unity from the predictions of GR at the present epoch.  (Weaker constraints yet result if
the redshift dependence of $\varpi (z)$ is allowed to vary; see Ref.~\cite{Paolo}.)
These are not very tight constraints.  As shown in Sec.~\ref{forecasts}, 
it seems likely that experiments already under
consideration will give us much tighter constraints on parametrized-post-Friedmannian 
departures from GR in the near future.  If, indeed, future constraints improve, we may need
to reconsider the assumption of homogeneous $\varpi$.

%\cite{Koivisto:2005mm}
%\bibitem{Afshordi:2008rd}

Throughout this paper we neglect any possible scale-dependence of $\varpi$. This simplifying assumption seems justified given
the absence of any significant departure from GR.
Were we to see evidence of a departure from GR, the onus would be on us to demonstrate the new theory's consistency with solar system-scale tests, all of which prefer GR to one part in $10^5$ (e.g. Ref.~\cite{Bertotti:2003rm}).  Beyond this experimental evidence, we expect that $\varpi$ should be scale dependent simply due to the differing evolution histories of sub- and super-horizon perturbation modes.  Other work has already attempted tackling this expectation.  Hu and Sawicki implement a scale-dependent gravitational slip, based on the behavior seen in $f(R)$ models of gravity. \cite{Hu:2007pj}. Afshordi {\it et al}. offer a scale-dependent parametrization of $-(\psi-\phi)/(\phi+\psi)$ designed to be consistent with higher-dimensional generalizations of DGP gravity \cite{Afshordi:2008rd}.  Though they find that their parametrization is capable of describing effects qualitatively consistent with tensions in current data sets, none of those tensions is strong enough to warrant a detection of alternative gravity.  Koivisto and Mota explore a different set of new gravitational effects by supposing that dark energy is an imperfect (non-zero shear) fluid \cite{Koivisto:2005mm}.  Shear $\sigma$, like gravitational slip, affects the space-space, off-diagonal perturbed Einstein equation, $k^2(\phi-\psi)=12\pi G
a^2\bar\rho(1+w)\sigma$. The imperfect fluid introduces a dark flow, however, so that the gravitational effects are not fully equivalent to the results of gravitational slip. Like the present work, they find that data cannot yet definitively rule in or out the interesting regions of their parameter space. Specifically, they find that the effect of non-zero shear on the CMB anisotropy spectrum is weaker than the effect of $\varpi$ demonstrated in DCCM.

We have also shown that the modification of the Poisson equation follows uniquely from the
assumptions of our model: the enforced relationship between $\phi$ and $\psi$, stress-energy
conservation, and the absence of a preferred frame indicated by a ``dark flow''.  This must be
taken into account when conducting future tests of GR on cosmological scales.

%%%%%%%%%%%

\acknowledgments
This work  was supported by NSF CAREER AST-0349213 (RC) and AST-0645427 (AC).  AC and RC
thank Caltech for hospitality while this work was completed.  AM research is
supported by ASI contract I/016/07/0 ``COFIS''.

%%%%%%%%%%%%%%%%%%%%%%%%%%%%%%%%%%%%%%%%%%%%%%%%%%%%%%%%%%%%%%%%%%%%%%%%%%%%%%%%%%%%%%%%%%%%

%%%%%%%%%%%%%%%%%%%%%%%%%%%%%%%%%%%%%%%%%%%%%%%%%%%%%%%%%%%%%%%%%%%%%%%%%%%%%%%%%%%%%%%%%%%%  

\end{document}